\documentstyle[epsf]{article}

\begin{document}

\title{Three-body halos in two dimensions}

\author{E. Nielsen, D. V. Fedorov and A. S. Jensen\\
Institute of Physics and Astronomy,\\
Aarhus University, DK-8000 Aarhus C, Denmark}

\date{}
\maketitle

\begin{abstract}
A method to study weakly bound three-body quantum systems in two
dimensions is formulated in coordinate space for short-range
potentials.  Occurrences of spatially extended structures (halos) are
investigated. Borromean systems are shown to exist in two dimensions
for a certain class of potentials. An extensive numerical investigation
shows that a weakly bound two-body state gives rise to two weakly bound
three-body states, a reminiscence of the Efimov effect in three
dimensions.  The properties of these two states in the weak binding
limit turn out to be universal.

{PACS number(s): 03.65.Ge, 21.45.+v, 31.15.Ja, 02.60Nm}
\end{abstract}

\paragraph*{Introduction.} 
Characteristic properties of halo systems are binding energies much
smaller than the typical energy of the interaction and spatial
extensions much larger than the range of the potential
\cite{han95,fed93}. Three-body halos in {\it three dimensions} exhibit
many interesting features. Borromean systems, discovered now
experimentally, are bound three-body structures where none of the
two-body subsystems are bound \cite{zhu93,fed94,goy95}. The Thomas and
Efimov effects are anomalies due to a singularity which occurs in a
three-body system when the ratio of scattering length and effective
range is infinitely large \cite{tho35,fon79,mac86,fed93b,efi90}. Then
the effective three-body potential behaves as the inverse square of the
distance which results in infinitely many three-body bound states.

The Thomas effect corresponds to an infinitely small effective range.
The Efimov effect is associated with an infinitely large scattering
length where the wave functions of excited states reside in the tail
region of the effective three-body potential. Physical examples of
these might possibly exist in nature \cite{efi90,esr96,fed94b}.

The three-body halos in {\it two dimensions} are less studied and the
present investigations \cite{bru79,adh88,lim80,vugalter} are limited in
various ways.  It has been proven \cite{vugalter} that the number of
bound states is always finite.  However, the possibility of existence
of a Borromean state has not been examined and therefore the number of
bound states is not established.  Moreover the properties of these
states have not been investigated in detail.

In this letter we shall for simplicity only consider a system of {\it
three identical bosons} in {\it two dimensions}. The purpose is to
(i)~formulate an efficient method to solve the coordinate space Faddeev
equations in two dimensions, applicable for arbitrary short-range
two-body potentials, (ii)~derive asymptotic equations for the effective
three-body potential which alleviate crucially the numerical
investigations of weakly bound systems, (iii)~discuss the possible
existence of Borromean systems, halos, Efimov and Thomas effects and
(iv)~illustrate numerically the universal properties of the three-body
halos.

\paragraph*{The method.} 
We shall use the method developed for three dimensions \cite{fed93b}.
Let ${\bf r}_{jk}=({\bf r}_{j}-{\bf r}_{k})$ be the distance between
particles $j$ and $k$, ${\bf r}_{i(jk)}=-{\bf r}_{i}+({\bf r}_{j}+{\bf
r}_{k})/2$ the distance between particle $i$ and the center of mass of
particles $j$ and $k$. All particles have the same mass $m$. The Jacobi
coordinates are then introduced as ${\bf x}_{i}={\bf r}_{jk}/\sqrt{2}$,
${\bf y}_{i}={\bf r}_{i(jk)}\sqrt{2/3}$. The hyperspherical coordinates
in two dimensions are given by $\{ \rho, \Omega _{i}\} \equiv \{\rho
,\alpha_i ,\theta _{xi},\theta _{yi}\}$, $\rho =\sqrt{
x_{i}^{2}+y_{i}^{2}}$, $\alpha _{i}=\arctan (x_{i}/y_{i})$, $\theta
_{xi}$ and $\theta _{yi}$ are the azimuthal angles of ${\bf x}_{i}$ and
${\bf y}_{i} $.  The volume element in hyperspherical coordinates is
$\rho^3 {\rm d}\rho {\rm d}\theta_{x i}{\rm d}\theta_{y i} \sin\alpha_i
\cos\alpha_i {\rm d}\alpha_i$ and the kinetic energy operator is
\begin{equation}
T=\frac{\hbar ^{2}}{2m}\left( -\rho ^{-3/2}\frac{\partial ^{2}}{\partial
\rho ^{2}}\rho ^{3/2}+\frac{3/4}{\rho ^{2}}+\frac{\hat{\Lambda}^{2}}{\rho
^{2}}\right) \; , 
\end{equation}
\begin{eqnarray}
\hat{\Lambda}^{2}=-{\frac{{\partial }^{2}}{{\partial }\alpha _{i}^{2}}}%
-2\cot (2\alpha _{i}){\frac{{\partial }}{{\partial }\alpha _{i}}}
 \\ \nonumber -{\frac{1}{{%
\sin ^{2}(\alpha _{i})}}}{\frac{{\partial }^{2}}{{\partial }\theta _{xi}^{2}%
}}-{\frac{1}{{\cos ^{2}(\alpha _{i})}}}{\frac{{\partial }^{2}}{{\partial }%
\theta _{yi}^{2}}}\;.
\end{eqnarray} 

The total wavefunction is now expanded in a complete set of
hyperangular functions
\begin{equation}
 \Psi (\rho ,\Omega )=\frac{1}{\rho ^{3/2}}\sum_{n=1}^{\infty}
 f_{n}(\rho )\Phi _{n}(\rho ,\Omega ) \; ,
\end{equation}
where $\Phi _{n}(\rho ,\Omega )$ for each $\rho$ are chosen as
eigenfunctions of the hyperangular part of the Schr\"{o}dinger
equation
\begin{equation} \label{sch}
\left( \hat{\Lambda}^{2}+\frac{2m}{\hbar ^{2}}\rho ^{2}\sum_{i=1}^{3}
V({\bf r}_{i})\right) \Phi
_{n}=\lambda _{n}(\rho )\Phi _{n}\ .
\end{equation}
Here $V$ is the two-body potential and the expansion coefficients
$f_{n}(\rho )$ satisfy the system of coupled equations
\begin{eqnarray} \label{rad}
\left( -\frac{\partial ^{2}}{\partial \rho ^{2}}+\frac{\lambda _{n}+3/4}{%
\rho ^{2}}-Q_{nn}-\frac{2mE}{\hbar ^{2}}\right) f_{n}(\rho )
 \\ \nonumber
=\sum_{n^{\prime
}\neq n}\left( Q_{nn^{\prime }}+2P_{nn^{\prime }}\frac{\partial }{\partial
\rho }\right) f_{n^{\prime }}(\rho )\ ,
\end{eqnarray} 
\begin{eqnarray} 
Q_{nn^{\prime }}(\rho )=\int d\Omega \ \Phi _{n}^{*}(\rho ,\Omega
) \frac{\partial^2 }{\partial \rho^2 }
\Phi _{n^{\prime }}(\rho ,\Omega ) \label{qva}   \\
 P_{nn^{\prime }}(\rho )=\int d\Omega
\ \Phi _{n}^{*}(\rho ,\Omega )\frac{\partial }{\partial \rho }\Phi
_{n^{\prime }}(\rho ,\Omega ) \; . 
\end{eqnarray} 

The wavefunction $\Phi _{n}$ is written as a sum of three components,
each expressed in the corresponding system of Jacobi coordinates
\begin{equation}
\Phi _{n}=\sum_{i=1}^{3}\phi _{n}^{(i)}(\rho ,\Omega _{i}).
\end{equation}
these components satisfy the three Faddeev equations
\begin{equation}
\hat{\Lambda}^{2}\phi _{n}^{(i)}+\frac{2m}{\hbar ^{2}}\rho ^{2}V(r_i) 
 \Phi _{n} =\lambda _{n}(\rho
)\phi _{n}^{(i)},\ i=1,2,3.  \label{ang}
\end{equation}
The physical solutions are equivalent to the solutions of the
Schr\"{o}dinger equation in Eq.(\ref{sch}) but these equations are
better suited for descriptions of subtle correlations.

The s-wave motion is responsible both for the long distance behavior
and the Thomas and Efimov effects. We therefore restrict to s-waves
where the wavefunction $\phi _{n}^{(i)}$ only depends on $\rho$ and
$\alpha_i$.  The Faddeev components in Eq.(\ref{ang}) must all be
expressed in the same set of Jacobi coordinates. This
amounts for s-waves to rewriting one set of coordinates in terms of
the other and a subsequent integration over the angular variables
$\theta _{x}$ and $\theta _{y}$, i.e. 
\begin{equation}
 \phi(\rho ,\alpha^{\prime }) \rightarrow
\frac{1}{2\pi }\int_{0}^{2\pi }\phi \left( \rho ,\alpha ^{\prime }(\alpha
,\beta )\right) d\beta ,
\end{equation}
where $\alpha ^{\prime }$ is given by $\alpha$ and the variable
$\beta$, describing the rotation between two sets of Jacobi
coordinates, given by
\begin{equation}
\sin ^{2}\alpha ^{\prime }=\frac{1}{4}\sin ^{2}\alpha +\frac{3}{4}\cos
^{2}\alpha \pm \frac{\sqrt{3}}{2}\sin \alpha \cos \alpha \cos \beta , \hfill
\end{equation}
where the choice of $\pm$ is independent of $\beta$.  For three
identical bosons ($\phi \equiv \phi _{n}^{(i)}$) the Faddeev equations
then reduce to the three identical equations
\begin{eqnarray}  \label{iden}
\left( -{\frac{{\partial }^{2}}{{\partial }\alpha ^{2}}}-2\cot (2\alpha ){%
\frac{{\partial }}{{\partial }\alpha }}-\lambda \right) \phi (\rho
,\alpha ) =  \\ \nonumber
  - \frac{2m}{\hbar ^{2}}\rho ^{2}V(\sqrt{2}\rho \sin \alpha )\left(
\phi (\rho ,\alpha )+\frac{1}{\pi }\int_{0}^{2\pi }\phi (\rho ,\alpha
^{\prime })\ d\beta \right) .
\end{eqnarray}

\paragraph*{Large-distance behavior.} We define an angle $\alpha_0$
such that $\left| 2m\rho ^{2}V(\sqrt{2}\rho \sin \alpha_0)/\hbar
^{2}\right| =\left| \lambda(\rho)\right| $ and we assume that the
potential is of short range, that is $\alpha_0$ approaches zero with
increasing $\rho$.  If $\alpha > \alpha _{0}$ the potential
is then negligible for large $\rho$ and Eq.(\ref{iden})
becomes
\begin{equation}
\left( -{\frac{{\partial }^{2}}{{\partial }\alpha ^{2}}}-2\cot (2\alpha ){%
\frac{{\partial }}{{\partial }\alpha }}-\lambda \right) \phi (\alpha )=0.
\end{equation}
The solution which satisfies the boundary condition of vanishing derivative
at $\alpha =\pi /2$ is 
\begin{equation}
\phi (\alpha )=\pi \cos \pi \nu ~P_{\nu }(\cos 2\alpha )-2\sin \pi \nu
Q_{\nu }(\cos 2\alpha ),  \label{free}
\end{equation}
where $\lambda \equiv 4\nu (\nu +1)$ and $P$ and $Q$ are Legendre
functions.  At small $\alpha$ this solution behaves as
\begin{eqnarray} \label{bou}
\phi (\alpha )=2\sin (\nu \pi )\left( \gamma +\log \alpha +\psi (1+\nu
)\right)  \nonumber \\
 +\pi \cos (\nu \pi )+O(\alpha^{2}) \; ,
\end{eqnarray}
where $\psi$ is the digamma function and $\gamma$ is Euler's constant.
Without interactions the solution in Eq.(\ref{free}) with the boundary
condition of zero derivative at $\alpha =0$ provides the quantization
rule $\nu=0,1,2...$, see Eq.(\ref{bou}).

For a non-zero short-range potential the integral in Eq.(\ref{iden})
can be expanded for $\alpha <\alpha _{0} \ll 1$ as
\begin{equation}
\frac{1}{\pi }\int_{0}^{2\pi }\phi (\rho ,\alpha ^{\prime })\ d\beta =
 2 \phi (\frac{\pi }{3})+O(\alpha ^{2}).
\end{equation}
Then Eq.(\ref{iden}) simplifies to a differential equation with an
inhomogeneous term
\begin{eqnarray} \label{inh}
\Bigl( -{\frac{{\partial }^{2}}{{\partial }\alpha ^{2}}}-2\cot (2\alpha ){
\frac{{\partial }}{{\partial }\alpha }} + \frac{2m}{\hbar ^{2}}\rho ^{2}V(
\sqrt{2}\rho \sin \alpha )  \nonumber \\ 
   - \lambda \Bigr) \phi (\alpha )
  = -\frac{2m}{\hbar ^{2}}
\rho ^{2}V(\sqrt{2}\rho \sin \alpha )2\phi (\pi /3) \; .
\end{eqnarray}
The homogeneous part of this equation reduces for small $\alpha$ by the
substitution $r = \rho \alpha \sqrt{2}$ to the two dimensional
Schr\"{o}dinger equation for two particles. Thus, assuming that
$\lambda $ can be neglected compared to $2m\rho ^{2}V/\hbar ^{2}$,
the large-distance ($\alpha < \alpha _{0} \ll 1$) solutions to
the homogeneous part of Eq.(\ref{inh}) simply are the zero-energy
two-body solutions.  Then for $\alpha\simeq\alpha_0$ the physical
solution is approximately $C\log \left( \sqrt{2}\rho \alpha /a\right)
$, where $C$ is an arbitrary constant and $a$ is the scattering length
defined as the distance where the two-body wave function for zero
energy is zero. One solution to the inhomogeneous part of the equation
is now $-2 \phi (\pi /3)$. The complete
physical solution to the inhomogeneous equation is therefore ($\alpha
\simeq \alpha_0$)
\begin{equation}
\phi (\alpha )=C\log \left( \frac{\sqrt{2}\rho \alpha }{a}\right) 
-2\phi \left( \frac{\pi }{3}\right) .  \label{inhomo}
\end{equation}

Matching the two solutions,
Eqs.(\ref{free}) and (\ref{inhomo}), 
and their derivatives at $\alpha =\alpha _{0}$ gives the
equation
\begin{eqnarray} \label{eig}
2\sin (\nu \pi )\log \left( \frac{\sqrt{2}\rho }{a}\right) =2\sin (\nu \pi
)\left( \gamma +\psi (1+\nu )\right)  \nonumber \\
+\pi \cos (\nu \pi )+2\phi \left( 
\frac{\pi }{3}\right) .
\end{eqnarray}
Both for $\rho \gg a$ and $\rho \ll a$ the logarithm at the left hand
side is large. The quantity $\nu $ must therefore approach an integer
$l$ to compensate for this divergence. The leading order of an
expansion in powers of $\rho $ for the lowest $\nu$ gives
\begin{equation} \label{rholimits}
 \nu \approx  \frac{3}{2}
\left[ \log \left( \frac{4 \sqrt{2}\rho }{3 a}\right)\right]^{-1} 
\rightarrow 0  \; \; .
\end{equation}
Therefore this eigenvalue $\lambda \equiv 4\nu (\nu +1)$ is
approaching zero in both these limiting cases in contrast to three
dimensions where a negative constant asymptotically is approached when
the scattering length is infinitely large.  The effective radial
potential in Eq.(\ref{rad}) has therefore a repulsive centrifugal term
and no collapse of the wave function in the center is possible.

Let us now consider diverging solutions $\lambda(\rho)$ to
Eq.(\ref{eig}). Then large imaginary values of $\nu \propto -i\rho$ are
necessary and the outer function in Eq.(\ref{free}) approaches
\begin{equation} \label{out}
\phi (\alpha )\approx \frac{1}{\sqrt{\nu }}\sqrt{\frac{2\pi }{\sin
2\alpha }}\sin \left( (\nu +\frac{1}{2})(2\alpha -\pi )+\frac{\pi }{4}\right)
 \; ,
\end{equation}
which for large $\rho$ is exponentially small at $\alpha =\pi /3$
compared to it's value at $\alpha=\alpha_0$. For large imaginary $\nu $
the equation in Eq.(\ref{eig}) then becomes
\begin{equation} 
\log \left( \frac{\sqrt{2}\rho }{a}\right) =\gamma +\log \nu +\frac{1}{2\nu}
-\frac{1}{12\nu ^{2}}+i\frac{\pi }{2}  \; , 
\end{equation}
which has the asymptotic solution
\begin{eqnarray}\label{bound}
\nu &=&-\frac{1}{2}-ie^{-\gamma }\frac{\sqrt{2}\rho }{a}-\frac{i}{12}\left(
e^{-\gamma }\frac{\sqrt{2}\rho }{a}\right) ^{-1}+O(\rho ^{-2}) \\
\lambda &\equiv &4\nu (\nu +1)=-\frac{4}{3}-\frac{8}{e^{2\gamma }}\frac{\rho
^{2}}{a^{2}}+O(\rho ^{-2})  \label{lam} \; .
\end{eqnarray}
This parabolic behavior of $\lambda$ is the signature of a bound
two-body state with the binding energy $B$ and wave number $k$ given by
$4e^{-2\gamma }a^{-2}=2m^{*}B/\hbar ^{2}\equiv k^{2}$, where
$m^{*}=m/2$ is the reduced mass of the two particles. We can verify
this by solving the two-body problem, where the s-wave radial solution
outside the potential is $K_{0}(kr)\approx -\log{(kr/2)} - \gamma$
which must be matched with the solution inside the potential at
$r_0=\rho\alpha_0\sqrt{2}$.  For small binding we can use the
zero-energy solution which at $r\simeq r_0$ is $\log{(r/a)}$.  This
matching gives the above $B$ and $k$ and the result is accurate to the
order $r_0/a$.  For attractive zero-range potential it is therefore
exact.  For the potentials without two-body bound state, where $r_0/a$
is not small, this particular solution does not exist.

The wavefunction is exponentially small everywhere except in a small
region close to $\alpha =0$, see Eq.(\ref{out}).  In this region the
zero-range wave function obtained from Eq.(\ref{inh}) is proportional
to $K_{0}(kr=k\rho\alpha_0\sqrt{2})$ and after normalization
approximately given by
\begin{equation} 
\phi (\alpha )\approx 2k\rho K_{0}\left( \sqrt{2}k\rho \alpha \right) \; .
\end{equation}
The related diagonal part of $Q_{11}$ is then by use of Eq.(\ref{qva})
computed to be $Q_{11} = -\frac{1}{3}\frac{1}{\rho ^{2}}$, which in
combination with Eq.(\ref{lam}) gives the first diagonal equation in
Eq.(\ref{rad}) as
\begin{equation} 
\left( -\frac{\partial ^{2}}{\partial \rho ^{2}} - \frac{1}
{4\rho ^{2}} - \frac{2m}{\hbar ^{2}} (E+B) \right) f(\rho ) = 0 \; .
\end{equation}
This is the characteristic large-distance behavior of a two-body
radial Schr\"{o}dinger equation in two dimensions.

\paragraph*{Efimov and Thomas effects.}
The eigenvalue equation for large distances is in {\it three
dimensions} given by \cite{fed93b}
\begin{equation} \label{efim}
\sin (\widetilde{\nu }\frac{\pi }{2})\frac{\sqrt{2}\rho }{a}=-\widetilde{\nu 
}\cos (\widetilde{\nu }\frac{\pi }{2})+\frac{8}{\sqrt{3}}\sin (\widetilde{%
\nu }\frac{\pi }{6}),
\end{equation}
where $\lambda =\widetilde{\nu }^{2}-4$.  For $\rho \gg a$ the lowest
solution is $\widetilde{\nu }=2$ $(\lambda =0)$ that is a regular free
solution whereas for $\rho \ll a$, $\lambda \rightarrow \lambda
_{\infty }=-5.012...,$ which leads to a strongly attractive $\rho
^{-2}$ potential in the radial equation. Physically the $\rho ^{-2}$
behavior is limited at small distances by a finite range $R$ of the
potential, and at large distances by a finite scattering length $a$.
However, in the limit $R/a \rightarrow 0$ such a potential gives rise
to the so called ''falling towards the center'' phenomenon with
infinitely many bound states.  When $R\rightarrow 0$, the phenomenon is
called the Thomas effect and when $a \rightarrow \infty $ it
is called the Efimov effect.

\begin{figure}[t]
\centerline{\epsfbox[80 590 200 740]{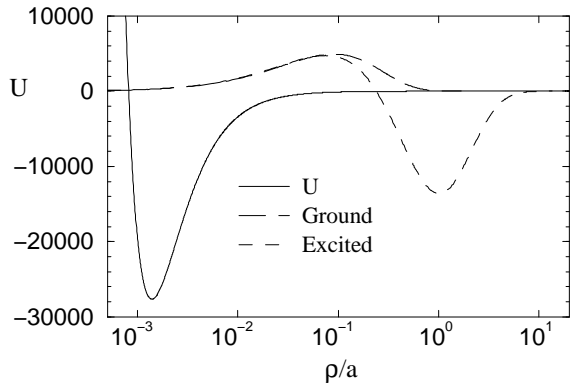}}
\caption{\small{
The effective radial potential,
$U=a^2\left[\left({3\over 4}+\lambda_1\right)\rho)^{-2}-Q_{11}\right]$,
as function of hyperradius (logarithmic scale) for a zero-range
two-body potential. The scattering length is $a$. The two unnormalized
bound-state wave functions are also plotted.
}}
\end{figure}

Eq.(\ref{efim}) is in two dimensions replaced by Eq.(\ref{eig})
and the Efimov effect is therefore not present in two
dimensions. Furthermore, the hyperradial potential has even for
zero-range potentials a repulsive centrifugal barrier at small
distance.  Then the Thomas collapse is not possible and the three-body
system must for zero-range potentials have a {\it finite} number
of bound states.

Equations (\ref{eig}) and (\ref{efim}) can be formally written in terms
of hypergeometric function as one general equation for $d$-dimensions,
where $d$ is a real number.  The asymp\-totic solution $\lambda_{\infty}$
of this equation leads to the Efimov and Thomas effects in the region
2.3$<d<$3.8.

\paragraph*{Bound states.}
The solutions to the zero-range potentials are similar to those of
purely attractive weak potentials.  The two-body system has at least
one bound state and for such interactions Borromean systems do not
exist in two dimensions. The three-body ground state is more bound and
an excited state is in addition always present.  In
Fig.1 is shown the effective radial potential
together with the resulting two bound-state wave functions.  The
corresponding root mean square radii are in units of the scattering
length respectively $<(r/a)^{2}>^{1/2} = 0.111, 0.927$.  Their large
sizes are reminiscent of the three-dimensional Efimov states
with an extension comparable to the scattering length.

The energies $E_3$ of these states are given in terms of the two-body
bound-state energy $E_2$ as $E_3/E_2 = 16.52, 1.267$. We have
numerically tried to find a third bound state with weaker binding
energy and larger radial extension by calculating the zero-energy
wavefunction and looking at the number of nodes. Even a careful search
to about 10$^3$ times the scattering length did not reveal another
bound state. This strongly indicates that only two bound states exist.
The large proportionality factor for the ground state energy can be
considered a reminiscence of the Thomas effect. These relations are
still valid for arbitrary weakly attractive, finite-range potentials.
This is because weak binding corresponds a scattering length much
larger than the range of the potential, which is the limit of a
zero-range potential.

In Fig.2 we show the ratio $(E_3-E_2)/E_2$ as function of $|E_2|$ for
different potentials. We first notice that there are always two
states which in the weak binding limit approach the results for
zero-range potential (marked with small circles). The purely attractive
potentials (solid lines) as well as repulsive core potentials (dashed
lines) fall approximately on the same universal curves. This is
equivalent to the observation in three dimensions that only low-energy
scattering properties like the scattering length are important in the
weak binding limit.

\begin{figure}[t]
\centerline{\epsfbox[80 590 200 740]{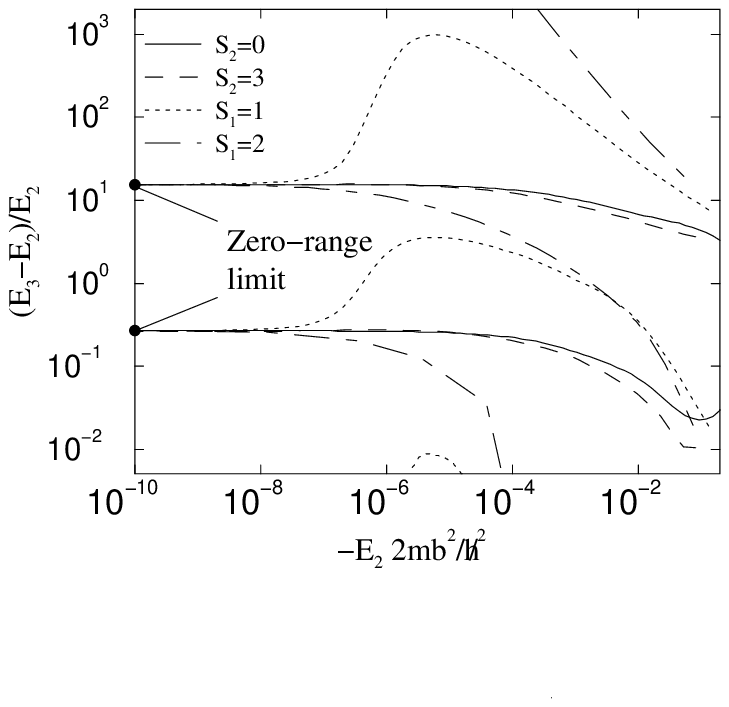}}
\caption{\small{
Ratio of three- to two-body energies as function of the
two-body energy for different two-body potentials $V(r)={\hbar^2\over
2mb^2} \left[S_1\exp{(-{1\over 2}r^2/b^2)}+
S_2\exp{(-2r^2/b^2)}\right]$. The unspecified strength parameter $S_i$
is used to vary the two-body binding.
}}
\end{figure}

The potentials with a short-range repulsive barrier (dash-dot and
dotted lines), unlike the repulsive core, produce energies deviating
in the middle of the plot from the universal curves.  The reason is that
a sufficiently large barrier and a sufficiently large attraction
produce a three-body ground state confined inside the barrier, that
yields three bound states in total. As $|E_2|$ then decreases towards
zero two cases are possible. If the potential does not allow the
spatially confined ground state the third bound state disappears and
the first two approach the zero-range limit (dotted lines).  If,
however, the potential allows a spatially confined ground state, all
three states survive in the small $|E_2|$ limit (dash-dot lines).  Now
the second and the third bound states approach the zero-range limit,
while the ground state energy remains finite. In this case the ground
state persists even into the region where two-body state is unbound,
creating thus a Borromean phenomenon.  This is only possible with a
repulsive confining barrier which therefore also limits the spatial
extension of the three-body system. The properties of Borromean systems
are therefore different in two and three dimensions.

\paragraph*{Conclusions.} Based on the hyperspherical expansion of the
Faddeev equations we have investigated the possible structure of three
weakly bound identical bosons in two dimensions.  For purely attractive
two-body potentials and the potentials with repulsive cores the
Borromean systems do not exist.  For two-body potentials with a
short-range repulsive barrier the two-body system may not have a bound
state while a three-body bound state exists. Borromean systems are
therefore possible in two dimensions.

We find numerically that a weakly bound two-body state is always
accompanied by two bound three-body states, resembling the Efimov
states in three dimension.  If a Borromean state is present there are
therefore in total three bound three-body states.  For all types of
potentials, two of these three-body bound states have energies and
radii following a universal curve in the weak binding limit. Their
sizes scale with the two-body scattering length and can therefore
become arbitrarily large in analogy to s-state halos in three
dimensions.


\begin{thebibliography}{99}
\bibitem{han95} P.G.~Hansen, A.S.~Jensen and B.~Jonson, 
 Ann.~Rev.~Nucl.~Part.~Sci. {\bf 45}, 591 (1995). 

\bibitem{fed93} D.V. Fedorov, A.S. Jensen, and K. Riisager, Phys.~Lett.
{\bf B312}, 1 (1993).

\bibitem{zhu93} M.V. Zhukov, B.V. Danilin, D.V. Fedorov, J.M. Bang,
I.J. Thompson and J.S. Vaagen, Phys.~Rep. {\bf 231}, 151 (1993).

\bibitem{fed94} D.V. Fedorov, A.S. Jensen, and K. Riisager, Phys.~Rev.
{\bf C49}, 201 (1994).

\bibitem{goy95} J. Goy, J.-M. Richard and S. Fleck, Phys.~Rev.
{\bf A52}, 3511 (1995).

\bibitem{tho35} L.H. Thomas, Phys.~Rev. {\bf 47}, 903 (1935).

\bibitem{fon79} A.C.~Fonseca, E.F.~Redish and P.E.~Shanley,
Nucl.~Phys. {\bf A320}, 273 (1979). 

\bibitem{mac86} J.H.~Macek, Z.~Phys. {\bf D3}, 31 (1986).

\bibitem{fed93b}  D.V.Fedorov and A.S.Jensen, 
Phys.~Rev.~Lett. {\bf 71}, 4103 (1993).

\bibitem{efi90} V.N. Efimov, 
Comm.~Nucl.~Part.~Phys. {\bf 19}, 271 (1990).

\bibitem{esr96} B.D. Esry, C.D.Lin and C.H. Greene, 
Phys.~Rev.~{\bf A54}, 394 (1996).

\bibitem{fed94b}D.V.~Fedorov, A.S.~Jensen and K.~Riisager, 
Phys.~Rev.~Lett.~{\bf 73}, 2817 (1994).

\bibitem{bru79}  L.W.Bruch and J.A.Tjon, Phys.Rev. {\bf A19}, 425 (1979);
J.A.Tjon, Phys.~Rev.~{\bf A21}, 1334 (1980).

\bibitem{adh88}  S.K.Adhikari, A.~Delfino, T.~Frederico, I.~D.~Goldman
and L.~Tomio, Phys.~Rev.~{\bf A37}, 3666 (1988).

\bibitem{lim80}  T.K.Lim and P.A.Maurone, Phys.Rev.~{\bf B22},1467 (1980).

\bibitem{vugalter}  S.~A.~Vugal'ter and G.~M.~Zhislin, 
Theor.~Mat.~Phys~{\bf 55}, 493 (1983).

\end{thebibliography}
\end{document}